\documentclass{ws-mpla}
\def\prl{{\em Phys. Rev. Lett. }}
\def\prc{{\em Phys. Rev. {\bf C} }}
\def\prd{{\em Phys. Rev. {\bf D} }}

\def\npa{{\em Nucl. Phys. {\bf A} }}
\def\npb{{\em Nucl. Phys. {\bf B} }}
\def\epjc{{\em Eur. Phys. J. {\bf C} }}

\def\plb{{\em Phys. Lett. {\bf B} }}

\def\pr{{\em Phys. Rep.}}

\def\app{{\em Acta Physica Polonica {\bf B} }}

\def\sup{{\em Supplemento al Nuovo Cimento} }
\def\num{{\em Nuovo Cimento} }

\def\EP{{ Europ.\ Phys.\ J.\ }}

\begin{document}
\markboth{J Cleymans and D Worku}{The Hagedorn Temperature Revisited.}
\title{The Hagedorn Temperature Revisited.}
\author{\footnotesize J. Cleymans}
\address{UCT-CERN Research Centre and Department of Physics, University of Cape Town, 
Rondebosch 7701, South Africa}
\author{\footnotesize D. Worku}
\address{UCT-CERN Research Centre and Department of Physics, University of Cape Town, 
Rondebosch 7701, South Africa}
\maketitle 
\begin{abstract}
The Hagedorn temperature, $T_H$ is determined from the number of hadronic resonances 
including all  mesons and baryons. This leads to a stable 
result $T_H = 174$ MeV consistent with the critical and the chemical freeze-out
temperatures at zero chemical potential. 
We use this result to  calculate the speed of sound and other thermodynamic quantities
in the resonance hadron gas model for a wide range of baryon chemical potentials
following the chemical freeze-out curve.
We compare some of our results to those obtained previously in other papers. 
\end{abstract}
\ccode{25.75,-q.05.70,-a,64.70,-p,64.90,+b}
\section{Introduction}
\begin{figure}[ht]
\begin{center}
\includegraphics[width=14cm, height=10cm]{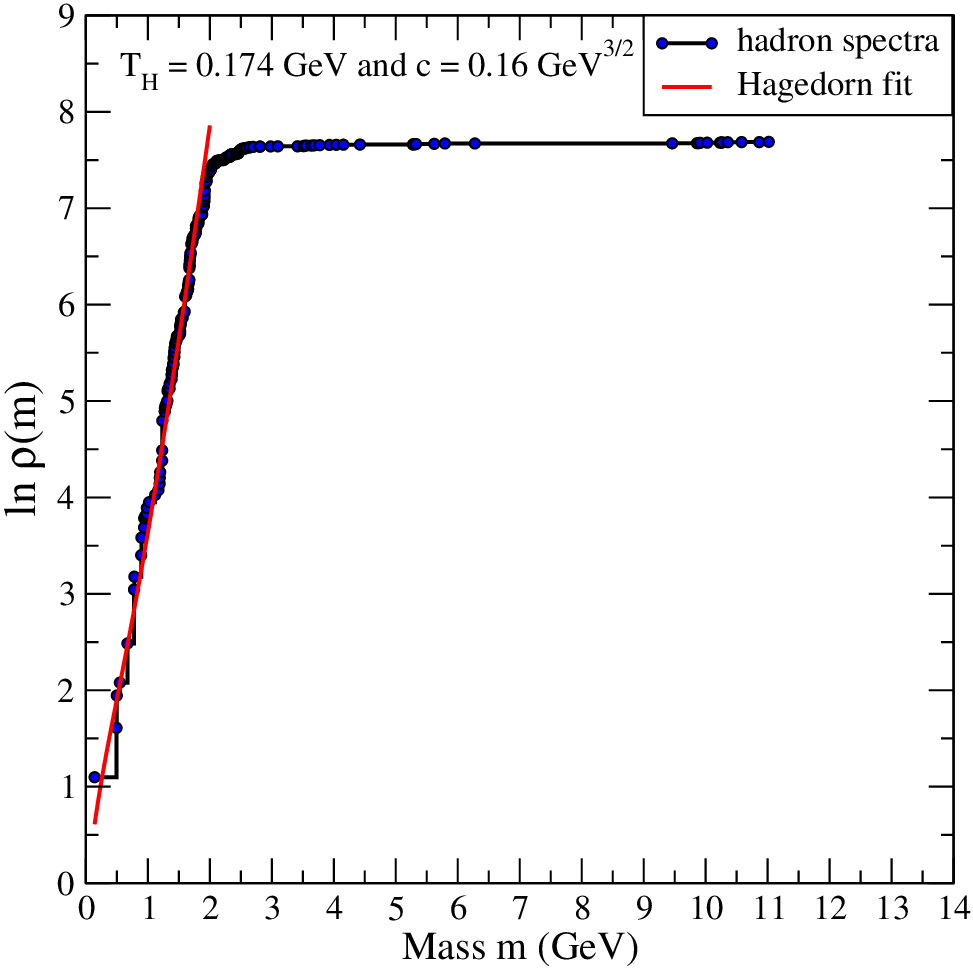}
\label{nocutoff}
\caption{Cumulative number of hadronic resonances as a function of $m$.
Again the hadronic data 
are made up of all resonances, including baryons, mesons and also heavy resonances
made up of charm and bottom quarks.}
\end{center}
\end{figure}

In 1965 Hagedorn ~\cite{hagedorn} proposed that the number of hadronic resonances
increases exponentially with the mass $m$ of the resonances. The idea, which was  
debated strongly  when first proposed, has since been widely accepted and  
 discussed by many authors~\cite{hagedorn-ranft,C-P,Vene,DRM,HW,S-ex,BFS,GG,KR-HS,KLP}.  
 The concept was based on 
the assumption that the  observed increase in the number of hadronic resonances would 
continue towards higher and higher masses as more experimental data became available~\cite{part}. 
The scale of the exponential 
increase determines the value of the Hagedorn temperature, $T_H$.
Recent papers ~\cite{MWT,MW,india,poland3,poland2,poland1} have used the latest results from 
the Particle Data Group~\cite{part}
to revisit the original analysis of Hagedorn to update the value of $T_H$.
This resulted in a surprising wide spread of possible values, with large variations as to whether
one considers mesons or baryons with values ranging from $T_H = 141$ MeV 
to $T_H = 340$ MeV
depending on the parametrization used  and on the set of hadrons (mesons or baryons). 
There thus exists  uncertainty as to the value of the Hagedorn temperature.
These have two origins:
\begin{itemize}
\item sparse  information about hadronic resonances certainly above 3 GeV, 
\item the analytical form of the Hagedorn spectrum, especially the factor multiplying the 
exponential.
\end{itemize}
The first item will probably never be resolved satisfactorily due to the 
width of resonances and also due to their large number making it difficult to 
identify them.
Splitting the spectrum into baryons and mesons further decreases the quality of the 
fits to the mass dependence of the 
mass spectrum.
We therefore propose to  stick to the 
original analysis of Hagedorn and consider a sum
over all resonances, baryons, mesons, strange, non-strange,  charm,  bottom etc.. 
This is the state that is produced at the Large Hadron Collider (LHC), namely, 
a hadronic 
ensemble containing all possible resonances.
The result is shown in Fig.~\ref{nocutoff} and leads to a  good determination 
of $T_H$ because the range 
in $m$ is reasonably large extending up 
to 3 GeV before reaching a plateau presumably due to the parsity of hadronic resonances 
above this value. Details about the parametrization used will be presented below.\\
The Hagedorn temperature  naturally leads to the notion that hadronic matter 
cannot exceed a limiting temperature 
 and increasing the beam  energy in proton$-$proton and 
proton$-$antiproton̄ collisions results in more and more hadronic resonances being produced
without a corresponding increase in the temperature of the final (freeze-out) state.  
This is the situation observed at 
the highest energies at the LHC.
It was  suggested long ago~\cite{C-P} that
the Hagedorn temperature is connected to the existence of a different phase in which quarks and gluons 
are no longer confined.
At present, the Hagedorn temperature is often understood as the temperature of the phase transition 
from hadrons to a quark-gluon plasma.\\
A recent analysis~\cite{india,And}  investigated the  
hadron resonance gas model~\cite{wheaton_phd,andronic,becattini} to show the stability of various thermodynamic quantities
in heavy-ion collisions when the Hagedorn spectrum is used literally, i.e. without a cut-off
on the number of resonances beyond a certain mass. In particular, 
the authors explored the addition of the 
hadron resonance gas model including an exponentially large  number of undiscovered resonances
which are naturally included in the Hagedorn model. 
Their results showed that the 
hadron resonance gas model gave different results 
for thermodynamic quantities but the overall chemical analysis was reasonably stable.

Furthermore, the use of so-called Hagedorn States (HS), based on the exponentially
increasing spectrum,  close to the  critical temperature can 
explain fast chemical equilibration by HS  
regeneration~\cite{Greiner} and  provide a unique method to compare lattice 
results for $T_c$ using thermal fits
HS provide a lower $\chi^2$ than thermal fits without HS \cite{Noronha}. 
These authors estimated effects by extending the hadron mass 
spectrum beyond $3$ GeV for $T_H$ = 200 MeV and assume that 
high mass excited $K^{*}$ 
states produce one kaon while  producing 
multiple pions thus  further reducing the $K^{+}/\pi^{+}$ \cite{And}. 

In this paper, we extend previous work~\cite{CRS,greiner2,Clyms} by 
extending explicit expressions of relevant 
thermodynamic quantities for non-zero chemical potentials. 
In particular the speed of sound, $C_{s}^{2}$, can be 
considered as a sensitive indicator of the critical behavior
 in strongly interacting matter. The results show a sharp 
dip of $C_{s}^{2}$ in the critical region, which is an 
indication that 
thermodynamics in the vicinity of confinement is indeed driven by the
higher excited hadronic states.\\ 
The outline of this paper is as follows. In section 2, we present the influence of the Hagedorn spectrum 
on the hadron yields to find the thermodynamic parameters and explain the basic 
concepts used in this paper. In section 3,  
we derive the number, energy and entropy densities and also the 
speed of sound. 
In section 4, we show the results using the  hadron and its extension resonance gas model 
for  particular thermodynamic quantities 
and discuss the relationship of $C_{s}^{2}$ with temperature and chemical potentials and compare
thermodynamic quantities, like energy and entropy density. 
The last section covers the conclusion. 

\section{Motivation}
 The particle data table contains hundreds of hadronic resonances~\cite{part} 
including the well known stable  hadrons like nucleons, pions, 
hyperons ($\omega$, $\Sigma$, $\Xi$), kaons etc..
In our calculations  we used 
a list of hadronic resonances including 
in total $250$ baryons and antibaryons, $348$ 
mesons and antimesons (counted without considering their 
isospin and spin degeneracies). 
Most of the hadronic resonances decay quickly  via strong interactions 
before reaching the
 detector, hence they are usually identified via their decay products. 
The mass of a decaying particle is equal 
to the total energy of the products measured in its rest frame.\\
 The basic idea of this paper is to add resonances using the  Hagedorn model for the spectrum. 
Using the hadronic data 
we show the relationship 
between the number of hadronic resonances and the  mass in
Fig.~\ref{nocutoff} where we took hadrons with masses up to $11.019$ GeV. 
The density of states obtained this way has been fitted using 
the following equation~\cite{hagedorn-ranft};
\begin{equation}\label{model2}
  \rho_{h}(m) = \frac{c}{\left(m^{2}+m_{0}^{2}\right)^{5/4}}\exp{\left(\frac{m}{T_H}\right)},
\end{equation}
where $c$ and $m_0$ are constant parameters given in the table below
\begin{table}[ht]
\begin{center}
\begin{tabular}{|c|c|}
\hline
Parameter             &                      \\
\hline 
\hline
$c$ $(GeV)^{3/2}$     & 0.16 $\pm$ 0.02      \\
$m_0$ $(GeV)$         & 0.5                  \\
$T_H$ $(GeV)$         & 0.174 $\pm$ 0.011    \\
\hline  
\end{tabular}
\caption{The result for the parameters of $c$, $m_0$ and $T_H$ using equation $(1)$.}
\end{center}
\end{table}
\section{The Hadron Resonance Gas Model and its Extension}
The thermodynamic properties  
of the Hadron Resonance Gas Model (HRGM) can be determined 
 from the partition function 
\begin{eqnarray}\nonumber \label{HRG}
\ln Z(V,T,\mu)& = &\int dm \left[\rho_{M}(m)\ln Z_{b}(m,V,T,\mu)\right.\\
&& \left. +\hspace{0.3cm}\rho_{B}(m)\ln Z_{f}(m,V,T,\mu)\right],
\end{eqnarray}
where the gas is contained in a volume $V$, has a temperature $T$ and 
chemical potential $\mu$, $Z_b$ and $Z_f$ are the partition 
functions for an ideal gas of bosons and 
fermions respectively with mass $m$, $\rho_{M}(m)$ and $\rho_{B}(m)$ are the 
spectral density of mesons and baryons. By using Eq.~\ref{HRG}, one can compute 
the number denisty $n$, 
energy density $\varepsilon$, entropy density $s$, pressure $P$, speed of 
sound $C_{s}^{2}$ and specific heat $C_v$.\\
hadron properties enter these models 
through $\rho _{M,B}$. The HRGM model takes the observed spectrum of 
hadrons up to some cutoff of mass $\lambda$, defined by
\begin{equation}
\rho_{M,B}(m) = \sum_{i}^{m_{i}\leq \lambda}g_{i}\delta(m - m_{i}),
\end{equation}
where $m_i$ are the masses of the known hadrons and $g_{i}$ the degeneracy 
factor. 
In order to explore the stability of results obtained using the HRGM, 
variant of these models is often used in which one takes the observed spectrum 
of states up to a certain cutoff of mass $\lambda$ and above this one 
includes an exponentially rising cumulative density of hadron states, which is 
defined in Eq.~\ref{model2}. 
This defines the Extended Hadron Resonance Gas Model (EHRGM).
The density of states  becomes 
\begin{equation}\label{hrgm}
\rho_{h}(m) = \sum_{i}^{m_{i}\leq \lambda}g_{i}\delta(m - m_{i}) 
+\frac{c}{(m^{2}+m_{0}^{2})^{5/4}}\exp{\left(\frac{m}{T_{H}}\right)}\theta(m-\lambda),
\end{equation}
where the model parameter $c$ and $T_H$ are fitted to data on the 
cumulative distribution of the sets of hadrons, $h$. Typically this model 
uses $m_{0} =$ $0.5$ GeV. 
Moreover, the parameters are determined from the  hadronic spectrum with 
masses up to $3$ GeV as shown in Fig.~\ref{nocutoff}. 
The results for $c$ and $T_{H}$ are given in Table 1.
\section{Derivation of thermodynamic quantities in EHRGM}
We are considering here the Boltzmann distribution for simplicity in order to 
present some of the results in a compact form. 
The partition function for a single particle
is given by 
\begin{equation}
\ln Z = \frac{gVTm^{2}}{2\pi^2}K_{2}\left(\frac{m}{T}\right) \exp{\left(\frac{\mu}{T}\right)},
\end{equation}
where $K_{2}$ is modified Bessel function. The meson mass distribution is 
taken to be  given by
\begin{equation}
\rho_{M,B}(m) = \sum_{i=1}^{m_{i} < \lambda}g_{i}^{M,B}\delta(m-m_{i}) + \frac{c}{(m^{2}+m_{0}^{2})^{5/4}}\exp{\left(\frac{m}{T_H}\right)}\theta(m-\lambda),
\end{equation}
%
\subsection{Derivation for the particle densities and pressure}
The particle density can be written as the sum of the two terms
\begin{equation}
n = n_{M} + n_{B},
\end{equation}
where $n_{M} = \ln Z_{M}/V$ and $n_{B} = \ln Z_{B}/V$ are number densities
 of mesons and baryons respectively. They are defined by
\begin{eqnarray}\nonumber
n_{M,B}& = &\frac{T}{2\pi^2}\sum_{i=1}^{m_{i} < \lambda}\exp{\left(\frac{\mu_{i}}{T}\right)}
\left[ g_{i}^{M,B}m_{i}^{2}K_{2}\left(\frac{m_i}{T}\right)\hspace{0.3cm}+\right.\\
&& \left. c\int_{m = \lambda}^{\infty}\frac{m^{2}}{(m^{2}
+m_{0}^{2})^{5/4}}\exp{\left(\frac{m}{T_{H}}\right)}K_{2}\left(\frac{m}{T}\right)dm\right],
\end{eqnarray}
where $\mu_{i} = S_{i}\mu_{S} + B_{i}\mu_{B} + Q_{i}\mu_{Q}$, for our case we 
consider an isospin symmetric system, where $\mu_{Q} =$ $0.0$ GeV. 
The pressure is given by
\begin{equation}
P = T\frac{\partial \ln Z_{M}}{\partial V} + T\frac{\partial \ln Z_{B}}{\partial V} = P_{M} + P_{B},
\end{equation}
where $P_{M} = Tn_{M}$ and $P_{B} = Tn_{B}$ are the pressure of 
mesons and baryons respectively. 
\subsection{Derivation for energy density}
The energy density is given by
\begin{equation}
\varepsilon_{M,B} = \frac{T^{2}}{V}\frac{\partial \ln Z_{M,B}}{\partial T},
\end{equation}
where $K_{1}$ is the modified Bessel function, $\varepsilon_{M}$ and $\varepsilon_{B}$ are energy densities of mesons and baryons respectively. 
\begin{equation}
\varepsilon_{M,B} = \sum_{i=1}^{m<\lambda}\left[
\frac{T^{2}}{2\pi^{2}}\exp{\left(\frac{\mu_{i}}{T}\right)}
\left(g_{i}^{M,B}m_{i}^{2}\left[3K_{2}\left(\frac{m_i}{T}\right)
+ \frac{m_i}{T}K_{1}\left(\frac{m_i}{T}\right)\right] + A_{H}\right)\right],
\end{equation}
where 
$$A_{H} = c\int_{m}^{\infty}\frac{m^{2}}{(m^{2}+m_{0}^{2})^{5/4}}\exp{\left(\frac{m}{T_{H}}\right)}K_{2}
(\frac{m}{T})dm.$$\\
From the above expressions, we can obtain the  entropy density
\begin{equation}
s = \frac{\varepsilon + P - n_{S}\mu_{S} - n_{B}\mu_{B}}{T},
\end{equation}
where $n_{S}$ and $n_{B}$ are the net  number densities for strange and 
baryonic particles  respectively
\begin{equation}
n_{S(B)} = \frac{T}{V}\frac{\partial \ln Z}{\partial \mu_{S(B)}}.
\end{equation}
\subsection{Derivation of the  speed of sound}
In hydrodynamic models the speed of sound plays an important 
role in the evolution of a system and is an ingredient in 
the understanding
of the effects of a  phase transition~\cite{CRS,greiner2,Clyms,Bilic}. 
It is well-known~\cite{landau} that the speed of sound has to 
be calculated at constant 
entropy per particle $(s/n)$. This makes the calculation more complicated
than the one at zero chemical potential where it is sufficient to 
keep the temperature fixed. In our extension we take the 
condition~\cite{landau} into account for non-zero baryon and 
strangeness chemical potentials, imposing overall strangeness zero. 
The squared speed is thus  calculated starting from
\begin{equation}
C_{s}^{2}(T,\mu) = \left(\frac{\partial P}{\partial \varepsilon}\right)_{s/n},
\end{equation}
The complete expression for the speed of sound can be rewritten as
\begin{equation}
C_{s}^{2}(T,\mu) = \frac{\left(\frac{\partial P}{\partial T}\right) + \left(\frac{\partial P}{\partial \mu_{S}}\right)\left(\frac{d\mu_{S}}{dT}\right)
+\left(\frac{\partial P}{\partial \mu_{B}}\right)\left(\frac{d\mu_{B}}{dT}\right)}{\left(\frac{\partial \varepsilon}{\partial T}\right) + 
\left(\frac{\partial \varepsilon}{\partial \mu_{S}}\right)\left(\frac{d\mu_{S}}{dT}\right) + 
\left(\frac{\partial \varepsilon}{\partial \mu_{B}}\right)\left(\frac{d\mu_{B}}{dT}\right)},
\end{equation}
where the derivatives $\frac{d\mu_{B}}{dT}$ 
and $\frac{d\mu_{S}}{dT}$ can be evaluated from two conditions. 
The first condition comes from  the requirement that
the ratio $(s/n)$ must be kept fixed, 
\begin{equation}\label{condition1}
\frac{s}{n} = 0,
\end{equation}
while the second condition 
follows from the conservation of strangeness
\begin{equation}\label{condition2}
n_{S} = n_{\bar S}. 
\end{equation}
The resulting expressions for
 $\frac{d\mu_{B}}{dT}$ 
and $\frac{d\mu_{S}}{dT}$ are presented in detail in the appendix.
\section{Results using HRGM and EHRGM}
\begin{figure}[ht]
\begin{center}
\includegraphics[width=5.2in]{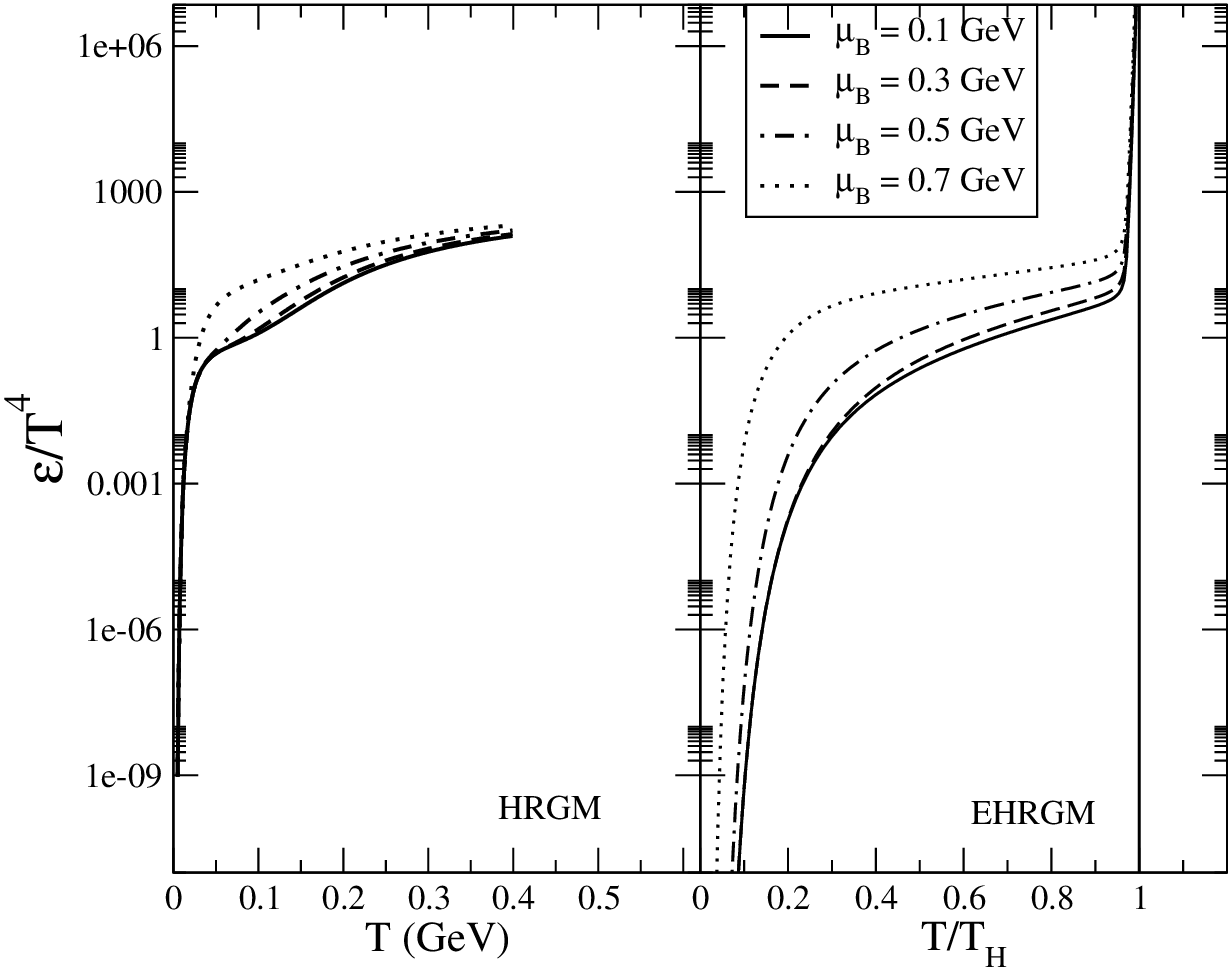}
\caption{(a) This figure show that the energy density $\varepsilon$ in units of $T^4$
calculated on the HRGM as a function of the temperature $T$.  
(b) This figure show that the energy density $\varepsilon$ in units of $T^4$ for EHRGM as a function of the $T$. 
The full$-$lines are the result of the EHRGM that accounts for all mesonic and baryonic resonances.}
\label{eperT4}
\end{center}
\end{figure}

\begin{figure}[ht]
\begin{center}
\includegraphics[width=5.2in]{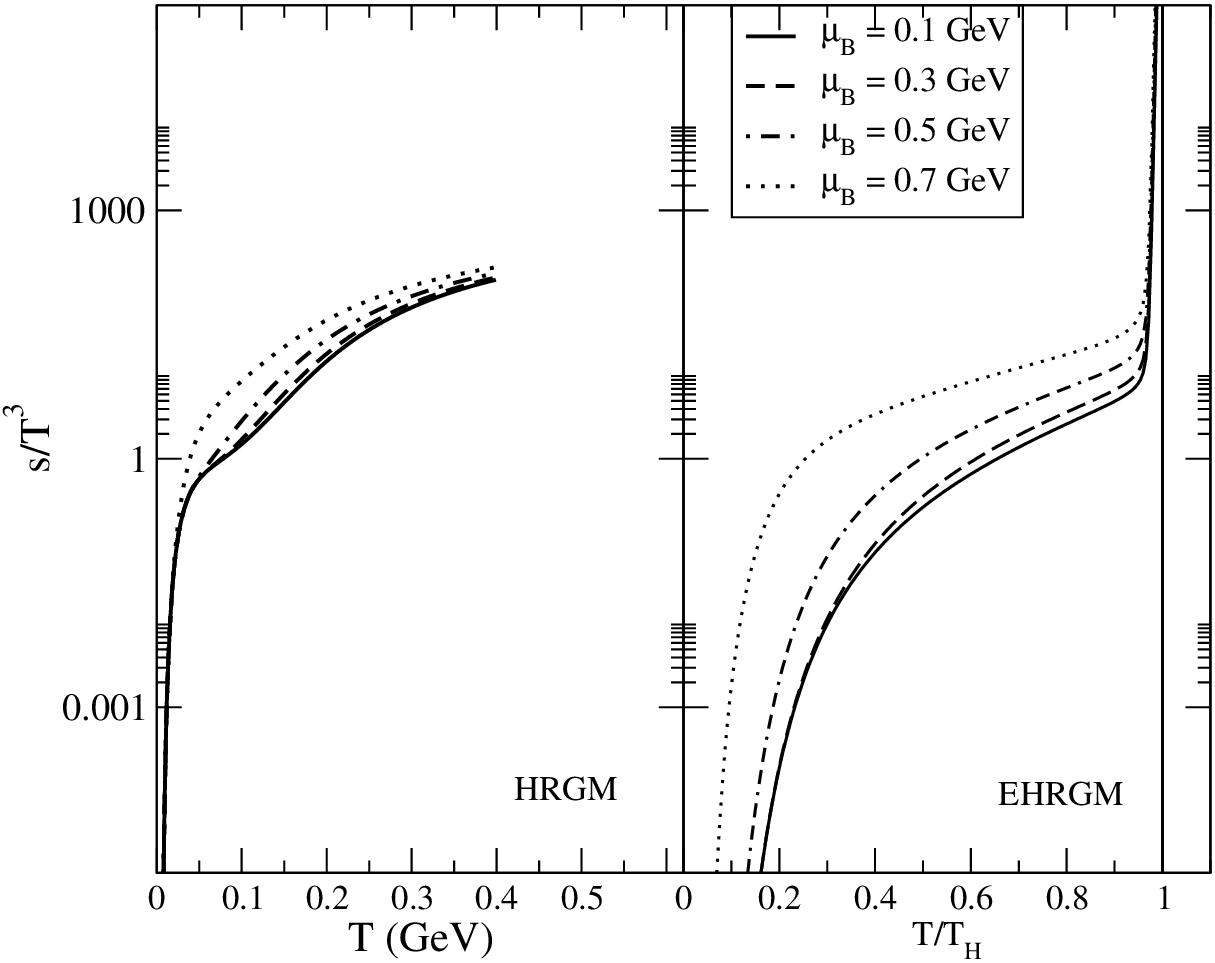}
\caption{(a)This figure show that the entropy $s$ in units of $T^3$
calculated in the HRGM as a function of the $T$.
(b) Entropy $s$ in units of $T^3$ using EHRGM as a function of the $T$. 
The full$-$lines are the result of the EHRGM that accounts for all 
mesonic and baryonic resonances.}
\label{SperT3}
\end{center}
\end{figure}

\begin{figure}[ht]
\begin{center}
\includegraphics[width=5.2in]{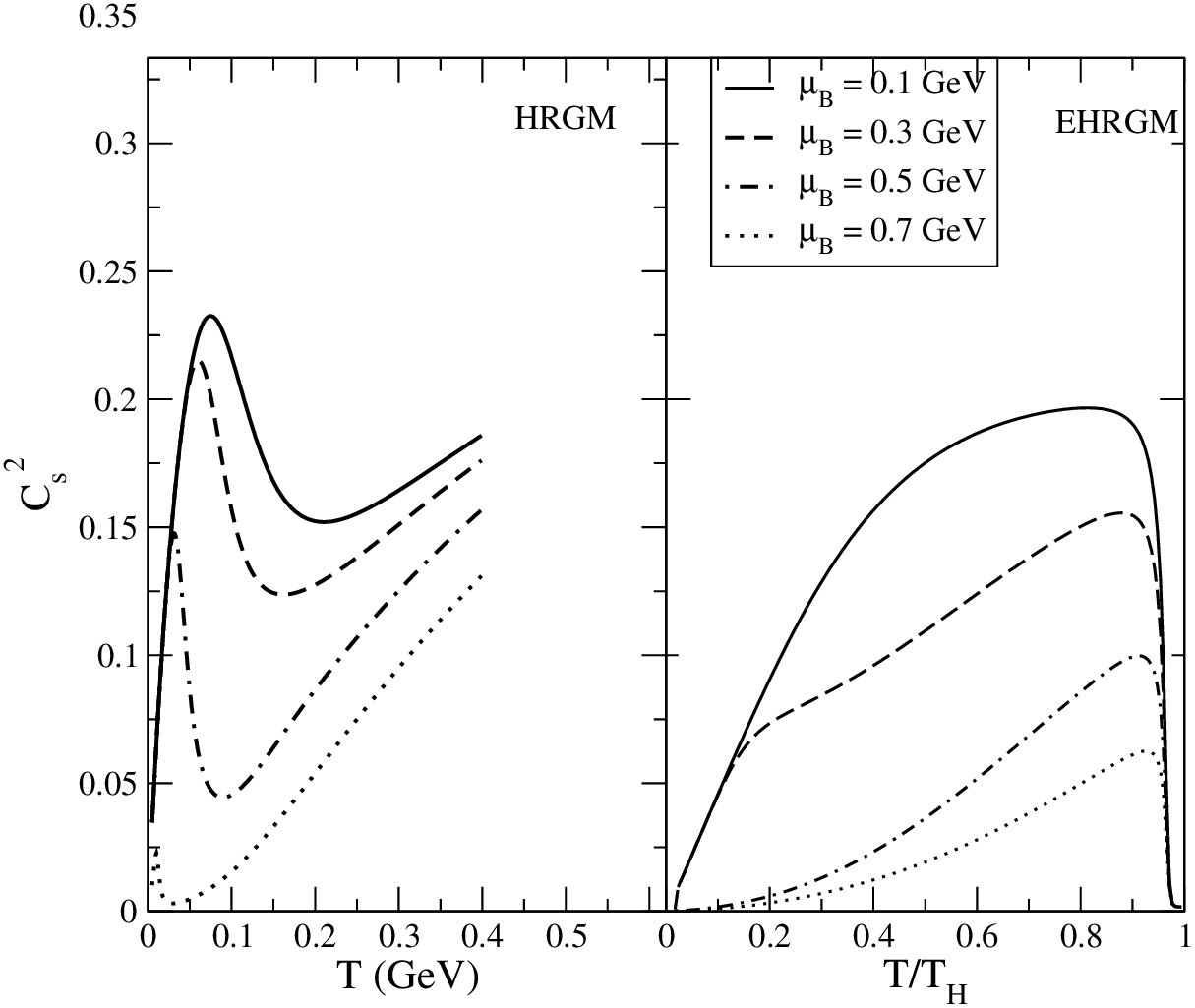}
\caption{(a) Showing the speed of sound $C^{2}_{s}(T,\mu)$ calculated on the HRGM as a function of the $T$ and $\mu$. 
(b) The speed of sound $C^{2}_{s}(T,\mu)$ in EHRGM with resonance mass 
truncation, $m < $ 2 GeV}
\label{CsHM}
\end{center}
\end{figure}

There is not much difference between the  HRGM and EHRGM models at low temperature, $T \ll T_H$ since the 
heavy resonances do not play an important role there.
Alternatively, at high temperatures, $T \gg T_H$, it 
should agree with the results of the lattice simulations of 
QCD \cite{fodor,hotqcd,physresgas}. In our case, we have determined 
the temperature and chemical potential dependencies of the pressure and the 
number, energy and entropy densities
using the chemical freeze-out curve~\cite{wheaton_phd,andronic,becattini}. 
Using these results, we have then calculated the speed of sound. 
The thermodynamic variables obtained using HRGM, we plot the energy and 
entropy densities scaled by the appropriate powers of $T$,
 which is shown in Fig.~\ref{eperT4}a, ~\ref{SperT3}a and ~\ref{CsHM}a, at 
the $T_{H} = 0.174$ GeV didn't observe any sudden change of the thermodynamic 
variables $\varepsilon/T^4$ and $s/T^3$ close to the critical temperature, it 
showed smooth shape as the temperature goes beyong $T_{H}$.\\
Moreover, when we use the EHRGM, we  see a different behaviour at the 
critical 
temperature which is shown in Fig.~\ref{eperT4}b, ~\ref{SperT3}b
and ~\ref{CsHM}b. These show that if the 
system undergoes a first-order phase transition, both temperature and pressure remain constant as hadronic matter is converted from 
hadron gas to quark-gluon plasma; the energy and entropy density, however, change 
discontinuously. This leads to show us a new state of matter.

The value of the speed of sound remains well below the ideal-gas limit 
for massless particles $C_{s}^{2} = 1/3$, even at very high 
temperatures, energy densities and pressure. 
In Fig.~\ref{CsHM}b represents, when it crosses the transition point 
in the case of full QCD,
we expect as before that $C_{s}^{2}$ should vanish 
beside that $C_{s}^{2}$ is inversely proportional to the 
specific heat $C_{v}$, and this 
can lead to  a divergence at the critical temperature $T_c$.
Of course, due to finite volume effects, the velocity of sound will 
 most likely not completely vanishes at $T = T_{c}$.  
The velocity of sound versus the temperature as shown in Fig.~\ref{CsHM}b is to evaluate $dP/d\varepsilon$, we have followed \cite{Bilic} and first expressed
$P$ and $\varepsilon$ in physical units, using EHRGM in Eq.~\ref{hrgm}. 
The temperature and chemical potential dependence of all the relevant thermodynamic quantities shows unique behaviour at the critical point
$T_c$, specially for the speed of sound, there is a pronounced dip as evidence for the phase transition in the system. Based on our calculations, 
the velocity of sound at $T = T_c$ is different from zero; it is expected to become zero but due to finite volume effects.
\section{Conclusions}
In this paper we have made a new analysis of the number of hadronic resonances
using the latest information from the Particle Data Group~\cite{part}.
This leads to a temperature which is consistent with the most recent results based on lattice QCD estimates of the phase transition 
temperature~\cite{fodor,hotqcd} and also the chemical freeze-out 
temperature 
at zero baryon density~\cite{wheaton_phd,andronic,becattini}.
We have extended calculations of the speed of sound to non-zero baryon
and strangeness chemical potentials keeping $s/n$ fixed~\cite{landau}. 
This is done for both the hadronic resonance gas model (HRGM) and the 
extended hadronic resonance gas model (EHRGM) which includes the 
exponentially increasing spectrum of hadrons following the Hagedorn 
parametrization.  The EHRGM shows that the speed of sound  goes to zero
at the phase transition point  while the HRGM shows a smooth dip 
followed by an increase.
\appendix
\section{Appendix: Speed of sound at non-zero chemical potentials.}
The speed of sound is given by~\cite{landau}
\begin{equation}
C_{s}^{2}(T,\mu) = \left(\frac{\partial P}{\partial \varepsilon}\right)_{\frac{s}{n}},
\end{equation}
where $s/n$ is the entropy per particle which is kept fixed.
using the variables $T, \mu_B$ and $\mu_S$, this can be rewritten as
\begin{equation}
C_{s}^{2}(T,\mu) = \frac{\left(\frac{\partial P}{\partial T}\right) + \left(\frac{\partial P}{\partial \mu_{S}}\right)\left(\frac{d\mu_{S}}{dT}\right)
+\left(\frac{\partial P}{\partial \mu_{B}}\right)\left(\frac{d\mu_{B}}{dT}\right)}{\left(\frac{\partial \varepsilon}{\partial T}\right) + 
\left(\frac{\partial \varepsilon}{\partial \mu_{S}}\right)\left(\frac{d\mu_{S}}{dT}\right) + 
\left(\frac{\partial \varepsilon}{\partial \mu_{B}}\right)\left(\frac{d\mu_{B}}{dT}\right)},
\end{equation}
where the derivatives $\frac{d\mu_{B}}{dT}$ and $\frac{d\mu_{S}}{dT}$ can be evaluated using two conditions. 
The first condition comes from 
keeping the ratio $(s/n)$ constant. From the  derivative one obtains
\begin{equation}
d\left(\frac{s}{n}\right) = 0,
\end{equation}
which implies
\begin{equation}
 nds = sdn.
\end{equation}
In terms of $T$, $\mu_B$ and $\mu_S$ this equation can be written as
\begin{eqnarray}
 n\left(\frac{\partial s}{\partial T}\right)dT &+&  
n\left(\frac{\partial s}{\partial \mu_B}\right)d\mu_{B} + n\left(\frac{\partial s}{\partial \mu_S}\right)d\mu_{S}\nonumber\\
&=&
 s\left(\frac{\partial n}{\partial T}\right)dT +
  s\left(\frac{\partial n}{\partial \mu_B}\right)d\mu_{B} + s\left(\frac{\partial n}{\partial \mu_S}\right)d\mu_{S}.
\end{eqnarray}
divide the above expression by $dT$ on both sides, so that it becomes
\begin{eqnarray}
 n\left(\frac{\partial s}{\partial T}\right) &+&  
n\left(\frac{\partial s}{\partial \mu_B}\right)\left(\frac{d\mu_{B}}{dT}\right) + n\left(\frac{\partial s}{\partial \mu_S}\right)\left(\frac{d\mu_{S}}{dT}\right)\nonumber\\
&=&
 s\left(\frac{\partial n}{\partial T}\right) +
  s\left(\frac{\partial n}{\partial \mu_B}\right)\left(\frac{d\mu_{B}}{dT}\right) + s\left(\frac{\partial n}{\partial \mu_S}\right)\left(\frac{d\mu_{S}}{dT}\right) .
\end{eqnarray}
Rearranging the above expression in order to write $\frac{d\mu_{B}}{dT}$ in terms of $\frac{d\mu_{S}}{dT}$ one obtains
\begin{eqnarray}
&& \left[n\left(\frac{\partial s}{\partial \mu_B}\right) -s\left(\frac{\partial n}{\partial \mu_B}\right)\right]\left(\frac{d\mu_{B}}{dT}\right)\nonumber\\
&&=
s\left(\frac{\partial n}{\partial T}\right) - n\left(\frac{\partial s}{\partial T}\right) - \left[n\left(\frac{\partial s}{\partial \mu_S}\right)
 -s\left(\frac{\partial n}{\partial \mu_S}\right)\right]\left(\frac{d\mu_{S}}{dT}\right).
\end{eqnarray}
Defining
\begin{eqnarray}
 A &=& n\left(\frac{\partial s}{\partial T}\right) - s\left(\frac{\partial n}{\partial T}\right),\\ 
 B &=& n\left(\frac{\partial s}{\partial \mu_{B}}\right) - s\left(\frac{\partial n}{\partial \mu_{B}}\right),\\
 C &=& n\left(\frac{\partial s}{\partial \mu_{S}}\right) - s\left(\frac{\partial n}{\partial \mu_{S}}\right). 
\end{eqnarray}
The final expression for condition one is
\begin{equation}\label{cond1}
 \frac{d\mu_B}{dT} = -\frac{1}{B}\left[A + C\frac{d\mu_S}{dT}\right].
\end{equation}
The second condition comes from overall strangeness neutrality, which is 
\begin{equation}\label{c1}
n_{S} = n_{\bar S} 
\end{equation}
where $n_S$ and $n_{\bar S}$ are the  strange and antistrange particle densities. 
Similarly the derivative of equation Eq.~\ref{c1} should thus satisfy 
\begin{equation}\label{c2}
 d(n_{S}) = d(n_{\bar S})
\end{equation}
this implies equation Eq.~\ref{c2} can be expressed as
\begin{eqnarray}
&& \left(\frac{\partial n_{S}}{\partial T}\right) +  
\left(\frac{\partial n_S}{\partial \mu_B}\right)\left(\frac{d\mu_{B}}{dT}\right) + \left(\frac{\partial n_S}{\partial \mu_S}\right)\left(\frac{d\mu_{S}}{dT}\right)\nonumber \\
&&=
 \left(\frac{\partial n_{\bar S}}{\partial T}\right) +
  \left(\frac{\partial n_{\bar S}}{\partial \mu_B}\right)\left(\frac{d\mu_{B}}{dT}\right) + \left(\frac{\partial n_{\bar S}}{\partial \mu_S}\right)\left(\frac{d\mu_{S}}{dT}\right).
\end{eqnarray}
We can apply the same method as above to write $\frac{d\mu_{B}}{dT}$ in terms 
of $\frac{d\mu_{S}}{dT}$ for the above relation
\begin{eqnarray}
&& \left[\left(\frac{\partial n_{S}}{\partial \mu_B}\right) -\left(\frac{\partial n_{\bar S}}{\partial \mu_B}\right)\right]\left(\frac{d\mu_{B}}{dT}\right)\nonumber\\
&=&
\left(\frac{\partial n_{\bar S}}{\partial T}\right) - \left(\frac{\partial n_{S}}{\partial T}\right) - \left[\left(\frac{\partial n_{S}}{\partial \mu_S}\right)
 -\left(\frac{\partial n_{\bar S}}{\partial \mu_S}\right)\right]\left(\frac{d\mu_{S}}{dT}\right),
\end{eqnarray}
we define $L = n_{S}$, hence it represents that the number of strangeness density for baryons and mesons
\begin{displaymath}
L = n_{S}^{B} + n_{S}^{M},
\end{displaymath}
and $R = n_{\bar S}$, the number of antistrangeness density for baryons and mesons
\begin{displaymath}
R = n_{S}^{\bar B} + n_{S}^{\bar M}.
\end{displaymath}
Define now
\begin{eqnarray}
E &=& \frac{\partial L}{\partial \mu_{B}} - \frac{\partial R}{\partial \mu_{B}}, \\
F &=& \frac{\partial L}{\partial \mu_{S}} - \frac{\partial R}{\partial \mu_{S}}, \\
D &=& \frac{\partial L}{\partial T} - \frac{\partial R}{\partial T}.
\end{eqnarray}
Hence, the final expression for condition two becomes
\begin{equation}\label{cond2}
 \frac{d\mu_B}{dT} = -\frac{1}{E}\left[D + F\frac{d\mu_S}{dT}\right].
\end{equation}
Finally, by equating equation ~\ref{cond1} and ~\ref{cond2} we find
\begin{equation}
\frac{d\mu_{S}}{dT} = \frac{AE - BD}{BF - CE},
\end{equation}
and
\begin{equation}
\frac{d\mu_{B}}{dT} = \frac{CD - AF}{BF - CE},
\end{equation}
Which is the relation used in the text.


\section*{References}
\begin{thebibliography}{90}
\bibitem{hagedorn} R. Hagedorn, \sup{} Volume III, 147 (1965); \num{35} (1965) 395; \num{56} A (1968) 1027.
%
\bibitem{hagedorn-ranft} R.~Hagedorn and J.~Ranft, \npb{48} (1972) 157-190.
%
\bibitem{C-P} N.~Cabibbo and G.~Parisi, \plb{59} (1975) 67.
%
\bibitem{Vene} G.\ Veneziano, \num{57A} (1968) 190.
%
\bibitem{DRM} K.~Bardakci and S.~Mandelstam, \pr{184} (1969) 1640;\\
S.~Fubini and G.~Veneziano, \num{64A} (1969) 811.

\bibitem{HW} K.\ Huang and S.\ Weinberg, \prl{25} (1970) 895.
%
\bibitem{S-ex} H.\ Satz, \prd{19} (1979) 1912.
%
\bibitem{BFS} Ph. Blanchard, S. Fortunato and H. Satz, \epjc{34} (2004) 361.
%
\bibitem{GG} R.\ V.\ Gavai and A.\ Goksch, \prd{33} (1986) 614.
%
\bibitem{KR-HS} K.\ Redlich and H.\ Satz, \prd{33} (1986) 3747. 
%
\bibitem{KLP} F.\ Karsch, E.\ Laermann and A.\ Peikert, \plb{478} (2000) 447.
\bibitem{part} Particle Data Group, C. Caso \textit{et al.},  \epjc{ 3}, (1998) 1-794 
\bibitem{MWT} M.\ Chojnacki, W.\ Florkowski and T.\ Csorgo, \prc{71} (2006) 044902.
\bibitem{MW} M.\ Chojnacki and W.\ Florkowski  \app{38} (2007) 3249.
\bibitem{india} S. Chatterjee, S. Gupta and R. M. Godbole, \prc{81} 044907 (2010).
\bibitem{poland3} W. Broniowski, W. Florkowski and L. Y. Glozman,  \prd{70} (2004) 117503
\bibitem{poland2} W. Broniowski,  {\it Preprint} hep-ph 0008112,  in ``Bled 2000: Few Quark Problems''.
\bibitem{poland1} W. Broniowski and W. Florkowski,  \plb{490} (2000) 223-227
\bibitem{wheaton_phd} J.\ Cleymans, H.\ Oeschler, K.\ Redlich, S.\ Wheaton, \prc{73} (2006) 034905.
\bibitem{andronic} A. Andronic, P. Braun-Munzinger, J. Stachel,  \npa{772} (2006) 167.
\bibitem{becattini} F.\ Becattini, J.\ Manninen, M.\ Gazdzicki, \prc{73} (2006) 044905.
%
\bibitem{Greiner} J. Noronha-Hostler, M. Beitel, C. Greiner, I. Shovkovy,  \prc{81} (2010) 054909.
%
\bibitem{Noronha} J. Noronha-Hostler, H. Ahmad, J. Noronha, C. Greiner, \prc{82} (2010) 024913.
%
\bibitem{And} A. Andronic, P. Braun-Munzinger, J. Stachel,  \plb{673} (2009) 142
\bibitem{CRS} P.\ Castorina, K.\ Redlich and H.\ Satz, \EP C 59 (2009) 67.
\bibitem{greiner2} J. Noronha-Hostler, J.~Noronha, C.~Greiner, \prl{103} (2009) 172302.
\bibitem{Clyms} P. Castorina, J. Cleymans,  D.E. Miller, H. Satz, \epjc{66} (2010) 207-213.\\
e-Print: arXiv:0906.2289 [hep-ph]
\bibitem{Bilic}J. Cleymans, N. Bilic, E. Suhonen and D.W. von Oertzen, \plb{311} (1993) 266-272. 
\bibitem{landau} L.D. Landau, E.M. Lifshitz (1987). 
Fluid Mechanics. Vol. 6 (2nd ed.). Butterworth-Heinemann. ISBN 978-0-080-33933-7.
%
\bibitem{fodor}  S.~Borsanyi {\it et al.} Wuppertal-Budapest Collaboration,
JHEP {\bf 1011} (2010) 077;
\bibitem{hotqcd}
 W.~S\"oldner (for the HotQCD collaboration) [arXiv:1012.4484] [hep-lat]].
%
\bibitem{physresgas} P.\ Braun-Munzinger and J.\ Stachel, 
\npa{606} (1996) 320;
D.\ Prorok and L.\ Turko, arXiv:hep-ph/0101220.


\end{thebibliography}
\end{document}